\documentclass[useAMS,usenatbib]{mnras}
\usepackage{times}
\usepackage{natbib}
\usepackage{amssymb,amsmath}
\usepackage{graphicx,color}
\usepackage{hyperref}
\usepackage[usenames,dvipsnames]{xcolor}
\usepackage[OT2,T1]{fontenc}
\usepackage[draft]{todonotes}
\usepackage{multirow}
\usepackage{mathtools}
\usepackage[normalem]{ulem}
\usepackage{lineno}
\usepackage{gensymb}
\usepackage{setspace}
\usepackage{tabularx}
\usepackage{soul}
\usepackage{bm}
\usepackage{ulem}
\title{Impact of Image Persistence in the Roman Space Telescope High-Latitude Survey}
\author[C.-H. Lin et al.]
{\parbox{\textwidth}{Chien-Hao Lin$^{1,2}$\thanks{E-mail: \texttt{chienhao.lin@duke.edu}},
Rachel Mandelbaum$^1$, M.~A. Troxel$^2$,\\ Christopher M. Hirata$^{3,4,5}$ and Mike Jarvis$^6$
}
\vspace{0.4cm}\
\\$^1$McWilliams Center for Cosmology, Department of Physics, 
Carnegie Mellon University, Pittsburgh, PA 15213, USA,
\\$^2$Department of Physics, Duke University, Durham, NC 27708, USA,
\\$^3$Center for Cosmology and Astroparticle Physics, The Ohio State University, 191 West Woodruff Avenue, Columbus, OH 43210, USA
\\$^4$Department of Physics, The Ohio State University, 191 West Woodruff Avenue, Columbus, OH 43210, USA
\\$^5$Department of Astronomy, The Ohio State University, 140 West 18th Avenue, Columbus, OH 43210, USA
\\$^6$Department of Physics and Astronomy, University of Pennsylvania, Philadelphia, PA 19104, USA
}
\begin{document}
\date{\today}
\maketitle

\begin{abstract}
The High Latitude Survey of the Nancy Grace Roman Space Telescope is expected to measure the positions and  shapes of hundreds of millions of galaxies in an area of 2220 deg$^2$. This survey will provide high-quality weak lensing data with unprecedented systematics control. The Roman Space Telescope will survey the sky in near infrared (NIR) bands using Teledyne H4RG HgCdTe photodiode arrays. These NIR arrays exhibit an effect called persistence: charges that are trapped in the photodiodes during earlier exposures are gradually released into later exposures, leading to contamination of the images and potentially to errors in measured galaxy properties such as fluxes and shapes. In this work, we use image simulations that incorporate the persistence effect to study its impact on galaxy shape measurements and weak lensing signals. No significant spatial correlations are found between the galaxy shape changes induced by persistence.  On the scales of interest for weak lensing cosmology, the effect of persistence on the weak lensing correlation function is about two orders of magnitude lower than the Roman Space Telescope additive shear error budget, indicating that the persistence effect is expected to be a subdominant contributor to the systematic error budget for weak lensing with the Roman Space Telescope given its current design.
\end{abstract}


\section{Introduction }

Weak gravitational lensing arises due to the deflection of light by the gravitational fields of the large-scale structure, leading to coherent galaxy shape (or `shear') distortions. Measuring the correlation functions of galaxy shapes is therefore a method by which we can measure the growth 
of structure in the Universe \citep[][]{2001PhR...340..291B,2007ApJS..172..239M, 2008ARNPS..58...99H, 2015RPPh...78h6901K, doi:10.1146/annurev-astro-081817-051928} and hence a powerful method for constraining cosmological parameters \citep[][]{heh14,2016ApJ...824...77J,  2021arXiv210513544S,2020PASJ...72...16H,2021A&A...645A.104A}.  

Since weak gravitational lensing shear is only a percent-level signal, weak lensing measurements rely on the use of large galaxy samples to reduce statistical uncertainties. In the upcoming Stage-IV surveys, including Euclid\footnote{\url{http://sci.esa.int/euclid/}} \citep{2011arXiv1110.3193L}, the Vera C.~Rubin Observatory Legacy Survey of Space and Time  \citep[LSST\footnote{\url{http://www.lsst.org/lsst}}:][]{2019ApJ...873..111I, 2009arXiv0912.0201L}, and the Nancy Grace Roman Space Telescope\footnote{\url{https://roman.gsfc.nasa.gov}} \citep{2015arXiv150303757S, 2019arXiv190205569A}, the statistical uncertainties of the weak lensing measurements are expected to reach sub-percent level precision. With such small statistical uncertainties, the future of weak lensing analysis requires a better understanding and more careful control of systematics to avoid systematic uncertainties dominating over statistical uncertainties.

In this work, we investigate a new source of weak lensing systematics due to the persistence effect of infrared detectors. The Roman Space Telescope will survey the sky in several near-infrared (NIR) bands using  H4RG HgCdTe photodiode arrays \citep{2014SPIE.9154E..2HP, 2020JATIS...6d6001M}, which have a larger format ($4088\times 4088$ active pixels) and a smaller physical pixel size (10 $\mu$m) than the previous generation of arrays \citep{2011ASPC..437..383B}. For these NIR arrays, charges that are trapped in the photodiodes during earlier exposure are gradually released into later exposures, leading to contamination of the images and (if uncorrected or imperfectly corrected) errors in  the measured properties of the objects in these images, including the galaxy shapes. In practice, a  correction scheme will be applied on the images to remove the persistence effect. However, since no correction is perfect, there will still be a residual persistence signal after applying the persistence correction.  This fact motivates carrying out a study of the impact of (uncorrected) persistence on galaxy shapes, to understand the magnitude of the effect on weak lensing.

Several past efforts have investigated the scientific impact of non-idealities in the HgCdTe detectors used by Roman \citep[e.g.,][]{2011PASP..123..179B, 2016PASP..128i5001K, 2017JInst..12C4009P, 2019arXiv191209481T}. For persistence, although some work has been done on modeling the image persistence effect and developing the correction scheme \citep{2012SPIE.8442E..1WL}, no published work so far has investigated the impact of image persistence in the context of weak lensing. For this work, we will use image simulations that incorporate the current best understanding of image persistence for the detectors to be used by Roman to study its impact on galaxy shapes and the weak lensing signal. The implementation of image persistence effect is available in the \textsc{GalSim} library, and is relevant to missions that uses NIR CMOS detectors (e.g. Hubble WFC3-IR and JWST).

In Section~\ref{sec:background}, we outline the background for this work: weak lensing, image persistence, and the model of persistence to be used in the simulations. In Section~\ref{sec:sim} we describe the  image simulations used for this work. We present our results in Section~\ref{sec:res} and conclude in Section~\ref{sec:con}.

\section{Background}
\label{sec:background}
In this section, we provide background on weak lensing cosmology and the persistence effect in NIR detectors.

\subsection{Weak lensing cosmic shear correlation function}

The weak lensing effect is mathematically approximated as a linear transformation that maps the unlensed location to the lensed location \citep{2015RPPh...78h6901K}. The transformation matrix $\mathcal{A}$, which connects the shape of a source with the observed images, can be written as 
\begin{equation}
\mathcal{A} = \left(\begin{array}{cc} 1-\kappa-\gamma_1 & -\gamma_2\\ -\gamma_2 & 1-\kappa+\gamma_1 \end{array}\right),
\end{equation}
where $\kappa$ is the convergence and $\gamma_1, \gamma_2$ are the two components of the spin-two shear $\gamma = \gamma_1 + \mathrm{i}\gamma_2$ in Cartesian coordinates. In weak lensing measurements, the change in the observed galaxy ellipticity $\Delta e$ is proportional to reduced shear $g$, defined as 
\begin{equation}
    \label{eq:reduced_shear}
    g = \frac{\gamma}{1-\kappa}.
\end{equation}
In the case of weak lensing, $|\kappa|, |\gamma| \ll 1$ is often assumed, allowing us to directly relate shear and galaxy ellipticities.

Conventionally, when considering shear correlations between pairs of galaxies, the coordinate system is rotated so that the galaxy separation vector is parallel to the $x$-axis. The shear components are then decomposed into the tangential direction ($\gamma_t$) and the cross direction ($\gamma_\times$) in the rotated coordinates:
\begin{equation}
\gamma_t = - \mathbf{Re}(\gamma e^{-2\mathrm{i}\phi}), \gamma_\times = - \mathbf{Im}(\gamma e^{-2\mathrm{i}\phi}),
\end{equation}
where $\phi$ is the polar angle of the separation vector of the two galaxies. 
From the shear components we can write the two shear correlation functions as a function of angular separation \citep{1991ApJ...380....1M}
\begin{equation}
\label{eq:xi}
\xi_\pm(\theta) = \left\langle \gamma_t\gamma_t \right\rangle(\theta) \pm \left\langle \gamma_\times\gamma_\times \right\rangle(\theta) .
\end{equation}
The cosmic shear correlation function encodes the effect induced by gravitational lensing. Since $\xi_{\pm}$ is directly related to the the redshift-dependent nonlinear matter power spectrum and lensing efficiency, the measurement of $\xi_{\pm}$ is used to constrain cosmological parameters. In this work, we quantify the impact of persistence on the estimated shears of galaxies in Roman images, and thereby calculate the impact on the correlation functions.

\subsection{Persistence Effect}
\label{sec:pers}

In this section, we describe the physical origin of and approximate parametric models for the persistence effect in NIR detectors.

\subsubsection{Theory of trapped charge carriers}

Persistence is effectively an increase in the dark current due to prior illuminations. For the near-infrared HgCdTe photodiode detector, the hypothesized mechanism of image persistence involves charge capture and charge emission in the $p-n$ junctions \citep{2008SPIE.7021E..0JS,  2012SPIE.8442E..4WR, 2014arXiv1402.4181A,2019JATIS...5c6004T}. 
The theory that image persistence arises from traps in the depletion region of $p-n$ junctions was first proposed by \citet{2008SPIE.7021E..0JS, 2008SPIE.7021E..0KS}. This trap theory has been widely accepted in the literature since then.
\cite{2012SPIE.8442E..1WL} provided an initial description of persistence in the Hubble Space Telescope WFC3-IR detector based on several calibration programs. Persistence is a strong
function of the fluence of the source in the original image. For pixels that were saturated, the current due to persistence exceeds the dark current for several exposures.

As the $p-n$ junctions are exposed to  illumination, the photo-generated charges, including electrons and holes, start to accumulate, leading to the reduction of the width of the depletion region.  Resetting the device to a resetting voltage recovers the depletion region, while some charges are still trapped in defects and are left behind. The trapped charges will be continuously released, and the delayed charge emission contaminate subsequent exposures. The magnitude of the persistence effect, usually quantified in terms of charges released per second (e$^-$/s), correlates strongly with the illumination level \citep{2012SPIE.8442E..1WL}. As the illumination approaches the saturation level of the detector, more defects are filled and this effect gets stronger.

We note that the signal actually measured in a NIR detector is related to the voltage across the $p-n$ junction; for charges that are trapped in the depletion zone and then released, the ``observed'' charge (i.e., change in signal times the usual conversion gain) is not necessarily equal to the true number of charges released. We quote all persistence models in terms of the observed effect on the signal, since this is both what is measured in persistence tests for {\slshape Roman} \citep{2020JATIS...6d6001M} and what is relevant for astronomical observations such as weak lensing.

\subsubsection{Persistence model}
\label{sec:pers_model}
Even though the hypothesized mechanism of charge capture is widely accepted to explain persistence, a proper physical theory to predict the behavior of persistence has not yet been developed. Instead, we currently rely on empirical models to describe the effect. An empirical exposure time dependent Fermi model was adopted to model the Hubble WFC3 IR curve \citep{2015wfc..rept...15L}. The persistence effect of H4RG detectors is approximately one order of magnitude weaker than that of WFC3 around the saturation level, but the Fermi model still serves as a good fitting law for HXRG persistence \citep{2020JATIS...6d6001M}. The exposure time dependent Fermi model follows the mathematical form:
\begin{equation}
    \label{eq:fermi}
    P(x,t) = A \left( \frac{1}{\exp\left[{\frac{-(x-x_0)}{\delta x}}\right]+1}\right) \left(\frac{x}{x_0}\right)^\alpha \left( \frac{t}{1000} \right)^{-\gamma },
\end{equation}

with illumination level $x$ (in unit of e$^-$), time after reset $t$ (second) and other fitting parameters. $P(x,t)$ in Eq.~\eqref{eq:fermi} is persistence ($e^- /s$) representing the released charge per second. The parameter $A$ is the overall normalization; $\alpha$ and $\gamma$ are the power law exponents. $x_0$ and $\delta x$ dominate the sharp increase of the Fermi function around saturation.

\begin{figure}
	\centering
	\includegraphics[width=\columnwidth]{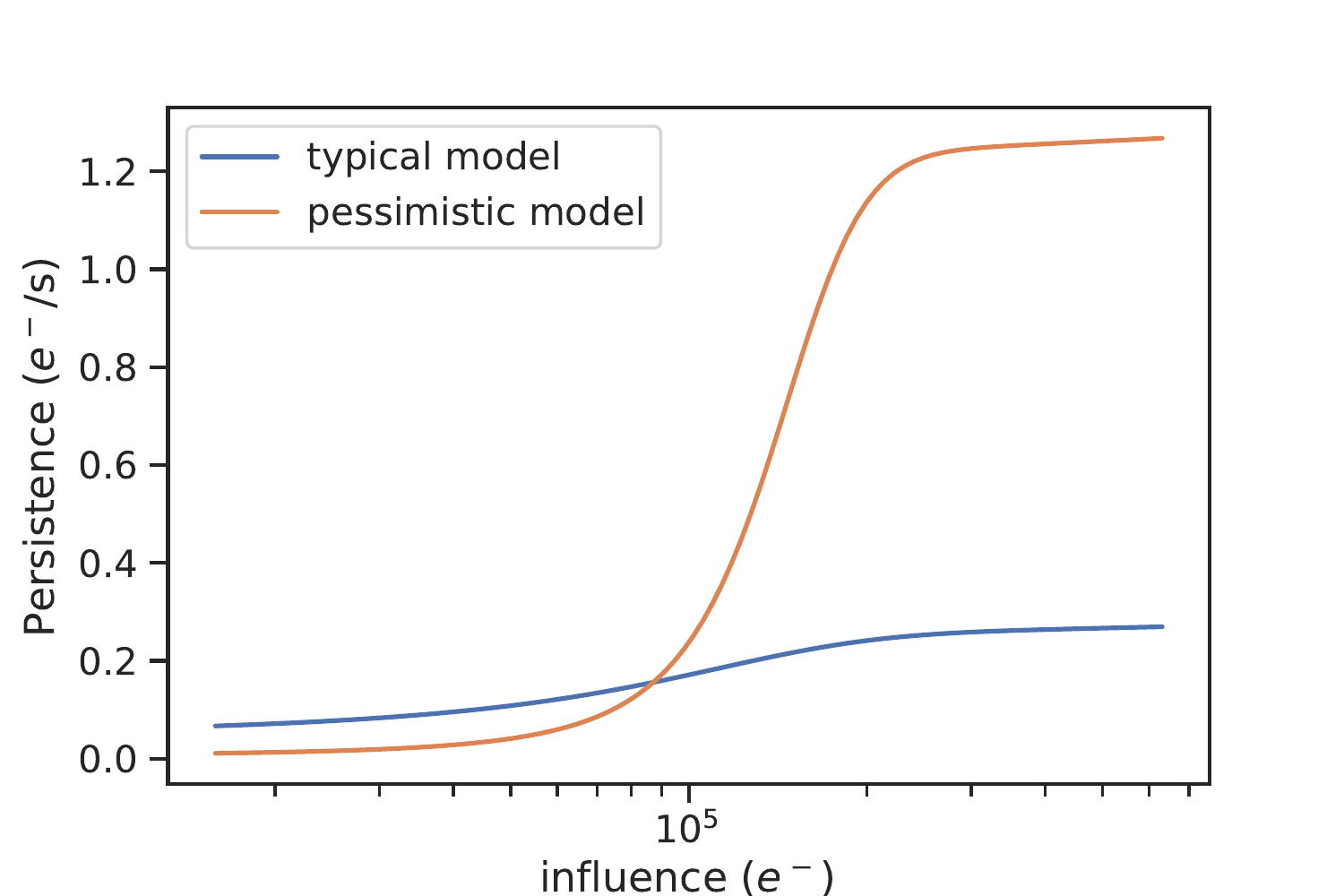}
	\caption{\label{fig:h4rg-lo}The curve of the H4RG persistence models with parameters specified in Table~\ref{tab:parameters}. The curves show the average of persistence current during the entire first 140.25 seconds after reset for various influence levels.
	}
\end{figure}

In this paper, we consider two sets of Fermi persistence model parameters for Roman detector candidates. The parameters in the Fermi model are fitted based on lab characterization of sample H4RG detectors\footnote{The parameters of the persistence characteristics are from the lab measurements of candidate detectors by the GSFC Detector Characterization Laboratory. Parameter values are obtained from internal communication with the Roman detector working group.}. Both sets of parameters are summarized in Table~\ref{tab:parameters} and the models are illustrated in Fig.~\ref{fig:h4rg-lo}. The 'typical model'  
is a more typical persistence model that depicts the persistence of most Roman detectors. There are, however, several detectors that exhibit strong persistence for high signals above 100k electrons. We consider both sets of model parameters, and will show how they impact the galaxy ellipticities in Sec.~\ref{sec:res}.

\begin{table}
\centering
\begin{tabular}{llllll}
\hline
&$A$   & $x_0$             & $\delta x$        & $\alpha$ & $\gamma$ \\ \hline
typical model &0.017 & $6.0 \times 10^4$ & $5.0 \times 10^4$ & 0.045    & 1        \\ \hline
pessimistic model &0.086 & $1.37 \times 10^5$ & $2.6 \times 10^4$ & 0.02    & 1        \\ \hline
\end{tabular}
\caption{\label{tab:parameters} The fitting parameters of the exposure time  dependent  Fermi  model as described in Eq.~\eqref{eq:fermi}.  }
\end{table}

\begin{figure*}
	\includegraphics[width=\textwidth]{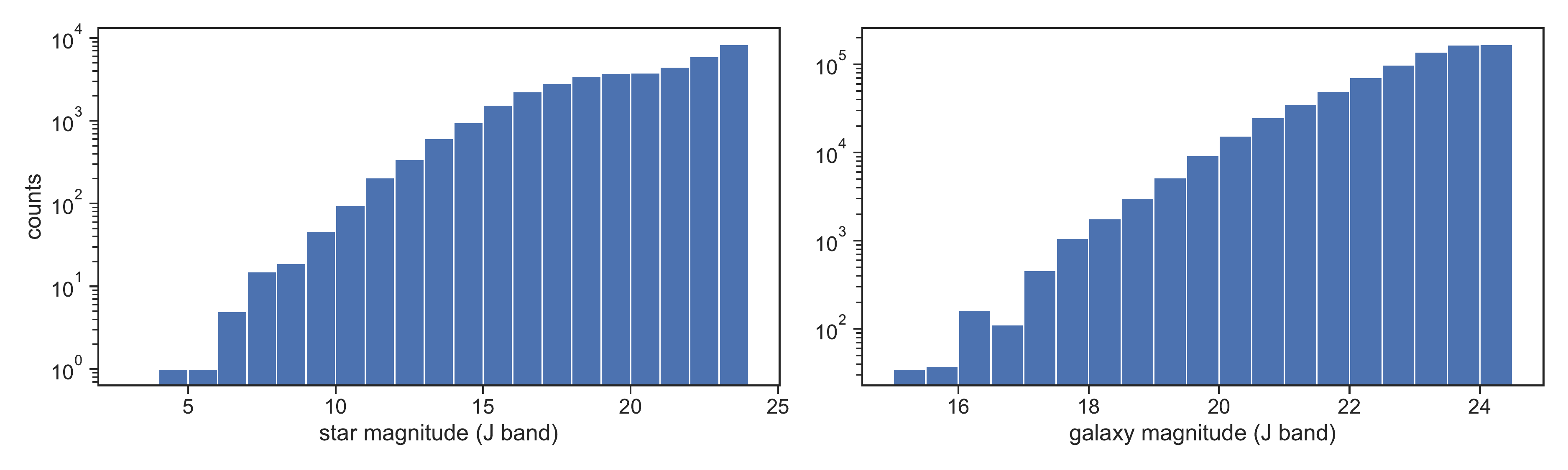}
	\caption{\label{fig:j_photometry}Magnitude distribution of simulated stars and galaxies in an area of 6.25 ${\rm deg}^2$. Left: the magnitude distribution of the simulated Galaxia stars in J band. Right: the simulated J-band magnitude distribution of the galaxy catalog. (Note that the x- and y-axis ranges of the two plots differ.) 
	}
\end{figure*}

\section{Image simulations}
\label{sec:sim}
In this work, we use the image simulation suite developed by \cite{2021MNRAS.501.2044T}. The persistence effect is applied along with other detector effects to test the impact of persistence on galaxy shapes. In practice, we focus on the persistence effect due to bright stars rather than galaxies, because bright stars that provide the flux level that results in strong persistence effects are far more numerous than bright galaxies at that flux level. 

Here we highlight the key elements of the simulations, and we refer readers to \cite{2021MNRAS.501.2044T} for more details.

\subsection{GalSim}
The image simulations in the work are carried out using the \textsc{GalSim}\footnote{\url{https://github.com/GalSim-developers/GalSim}} library for image rendering \citep{2015A&C....10..121R}. It can simulate galaxies from different galaxy models and also generate optical PSFs from parametric models. In particular, \textsc{GalSim} has a module especially designed for the image simulations for Roman Space Telescope \citep{2016PASP..128i5001K}. With \textsc{GalSim}, we simulate the images of star and galaxies for different bandpasses and Sensor Chip Assemblies (SCAs).

\subsection{Galaxy catalog}
The input galaxy catalog was generated by using random spatial locations in a $6.25~{\rm deg}^2$ sky area with a galaxy number density of $40~{\rm arcmin}^{-2}$. The photometric properties of the galaxies are drawn from a simulated Roman photometry catalog based on the Cosmic Assembly Near-infrared Deep Extragalactic Legacy Survey (CANDELS) survey\footnote{\url{https://github.com/WFIRST-HLS-Cosmology/Docs/wiki/Home-Wiki}} \citep{2019ApJ...877..117H}. In the right panel of Fig.~\ref{fig:j_photometry}, we illustrate the distribution of J-band magnitudes in the simulated photometry catalog.

\subsection{Star catalog}
The stars in the input catalog are simulated using Galaxia \citep{2011ApJ...730....3S}, a code that creates synthetic catalogs of stars in the Milky Way given magnitude limits and survey geometry. Galaxia in general accepts a variety of analytic models to simulate positions and ages of stars. In generating the star input catalog, we adopted the Besan\c con Milky Way model \citep{2003A&A...409..523R} thin and thick disk of the galaxy. The left panel of Fig. \ref{fig:j_photometry} shows the stellar magnitude distribution of the Galaxia simulated catalog, with star number density of $2.5~{\rm arcmin}^{-2}$.

\subsection{Implementation of persistence effect}
We implemented the Fermi model for the persistence effect in the Roman Space Telescope module of \textsc{GalSim}, as described in Sect.~\ref{sec:pers_model}.  For pixels with small fluence below half saturation well, the persistence signal increases linearly with fluence \citep{2019JATIS...5c6004T, 2020JATIS...6d6001M}, so we choose a linear model instead of the Fermi model due to lack of complete measurement of persistence at low fluence.

\subsection{Simulation steps}

\subsubsection{Building truth catalogs}
Before images are simulated, truth catalogs of objects to be simulated, including stars and galaxies, are constructed. The positions and magnitudes of stars follow the input catalog. For the galaxy truth catalog, besides the position and photometric properties, each galaxy is assigned an intrinsic ellipticity with ellipticity components drawn from a Gaussian distribution $G(0, 0.27)$ with a choice of typical intrinsic ellipticity dispersion \citep{2014MNRAS.439.1909V} and truncated at $0.7$ such that $(e_1^{\rm int})^2 + (e_2^{\rm int})^2 \le 1$. Since we are trying to identify whether there are additional coherent shear effects due to persistence rather than trying to recover the weak lensing shear signal, no coherent gravitational lensing shear is applied on top of the random intrinsic ellipticities in the galaxy catalog.

\subsubsection{Image simulation process}
The image simulation is carried out using the simulation suite developed by \cite{2021MNRAS.501.2044T}, which simulates the stars and galaxies in the input catalog and generates SCA images on $4088 \times 4088$ pixels stamps.  Effects including sky background, reciprocity failure, electron quantization, dark current, nonlinearity,  interpixel capacitance and readout noise are included in the image generation \citep{2021MNRAS.501.2044T}, following the order that each effect physically occurs. Besides these effect, we apply persistence from previous exposures using the model introduced in Sect. \ref{sec:pers_model}. Since the persistence decays with the resetting time, in the simulation process we limit the number of past exposures considered for persistence to 10 to facilitate the simulation process. For each galaxy, images with and without the persistence contamination are simulated to precisely measure the impact of persistence.

\subsubsection{Galaxy shape measurement}
Shape measurement is carried out by fitting the galaxy light profile to an exponential model using the Guassian mixture fitting module  \textsc{ngmix}\footnote{\url{https://github.com/esheldon/ngmix}} \citep{2015ascl.soft08008S} 
with 6 fitting parameters, including the centroid positions $x,y$, the ellipticities $e_1, e_2$, the size and the flux. Measurements are made on both simulations (without and with persistence included). The impact on galaxy shapes due to persistence is then captured in the difference in ellipticities for the same galaxy in the two sets of simulations, defined as 
\begin{equation}
    \Delta e = e_{\rm persistence} - e_{\rm intrinsic} = (\Delta e_1, \Delta e_2).
\end{equation}

It is worth noting that the intrinsic shape uncertainty of galaxies is removed when taking the the difference. This allows us to identify systematic biases in galaxy shapes due to persistence in a small area simulation. 

\section{Results}
\label{sec:res}
\subsection{Persistence effect in simulations}
\begin{figure}
	\centering
	\includegraphics[width=\columnwidth]{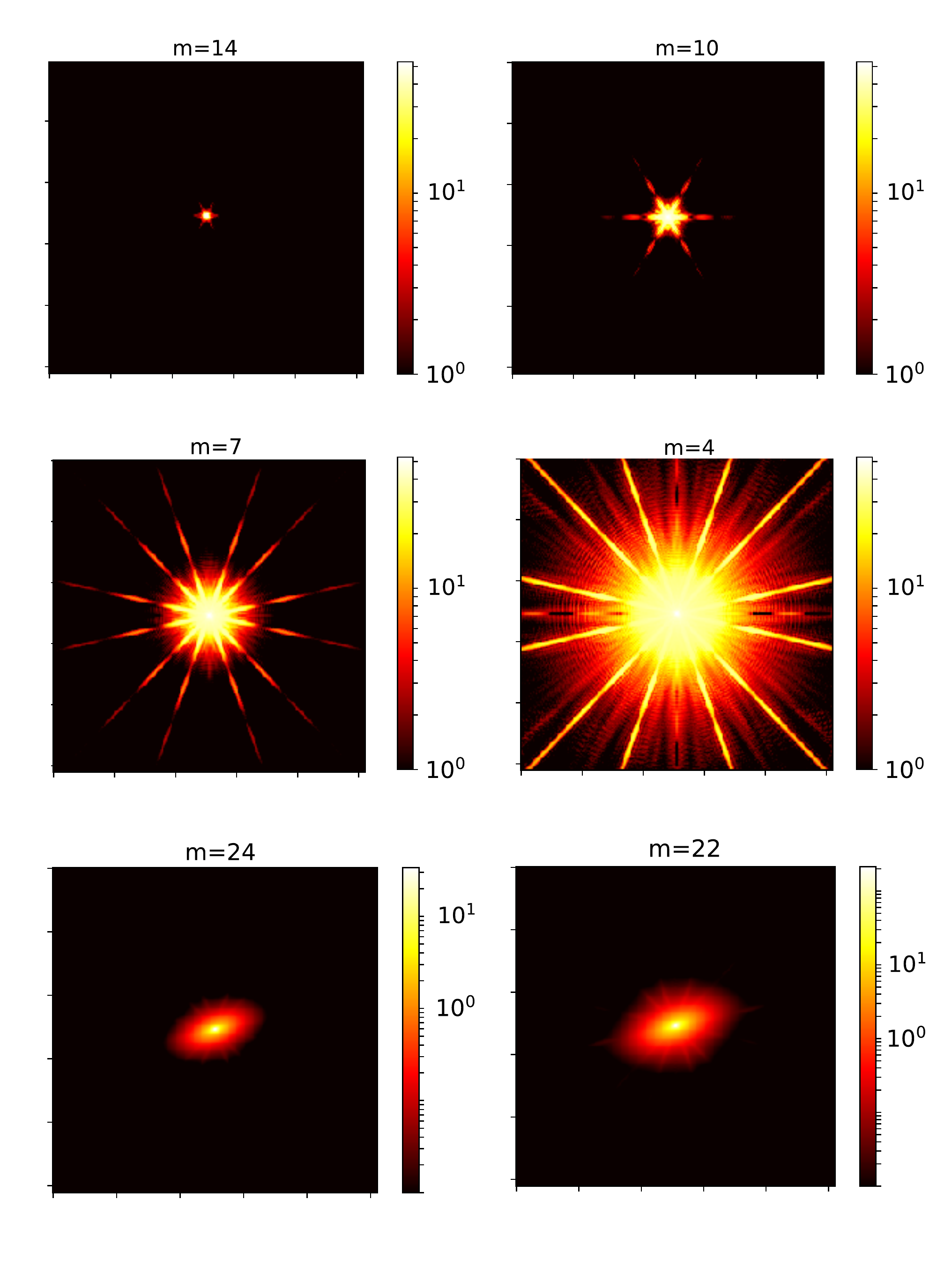}
	\caption{\label{fig:pers_joint} Top two rows: images of the persistence effect of stars with different magnitudes taken in 140s exposures after reset. Bottom row: simulated images of $J=22$ and $J=24$ galaxies captured by Roman in 140s exposures. The images are drawn on 256x256 images stamps in the Roman pixel scale. The color scales are in units of electrons with color maps on the right of each panel, but readers should note that the color bars for the star persistence (top two rows) and galaxies (last row) are not on the same scale. 
	}
\end{figure}

\begin{figure}
	\centering
	\includegraphics[width=\columnwidth]{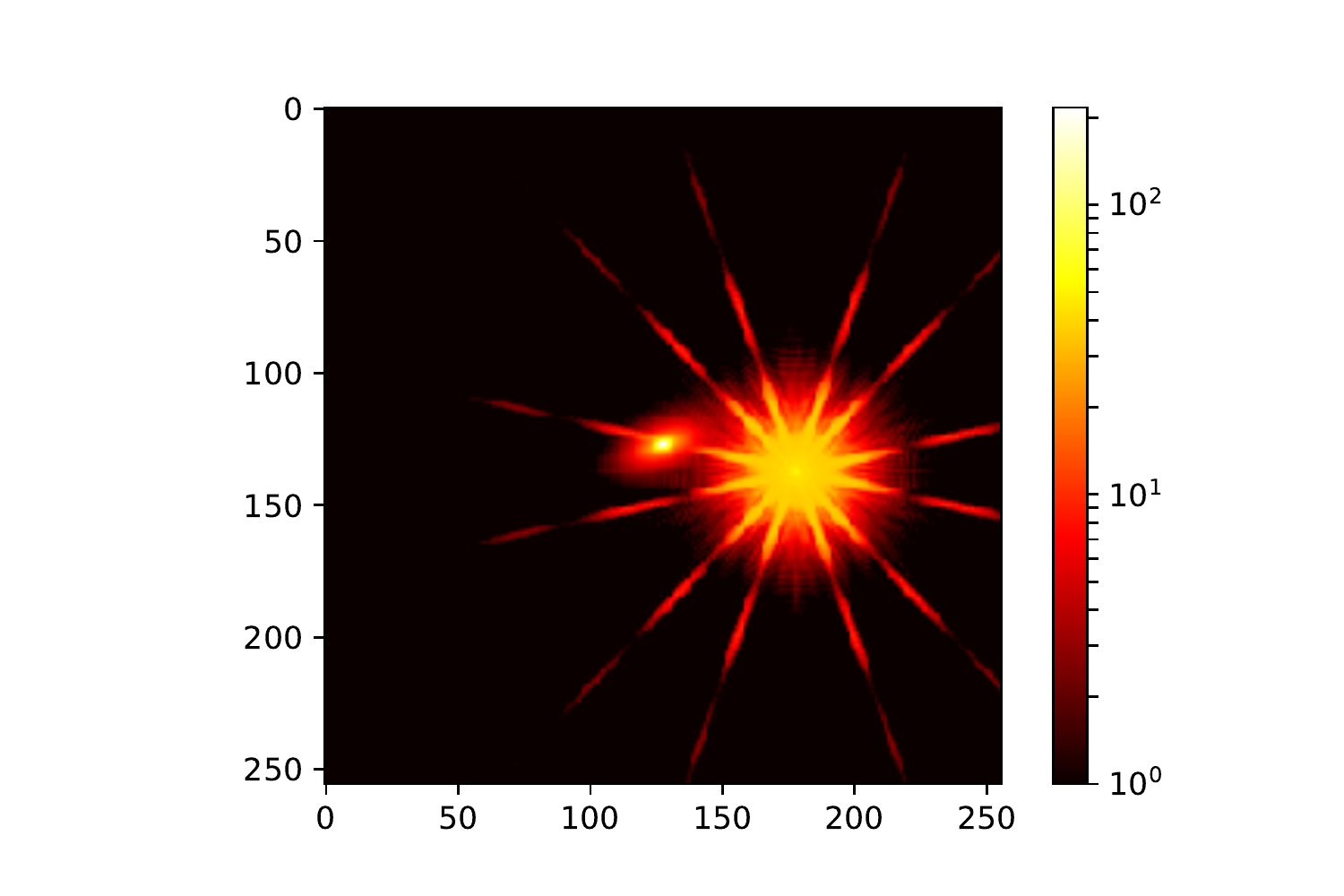}
	\caption{\label{fig:pers+gal} Illustration of the impact of star persistence on galaxy shapes in a 256x256 pixel image (x and y axes are pixel numbers). In this figure, a magnitude $J=22$ galaxy and the persistence of a magnitude $J=7$ star are plotted. In this case, the centroids of the two objects are located 51 pixels away from each other to illustrate the relative flux and size of the two components. 
	}
\end{figure}

To build intuition about how persistence effects from bright stars may affect the shapes of galaxies in subsequent exposures, we start by showing the image of the persistence effect due to bright stars. Fig.~\ref{fig:pers_joint} shows the persistence effect due to bright stars with a range of magnitudes; the effect is shown 140s after reset using the first (`typical') set of Fermi persistence model parameters specified in Tab.~\ref{tab:parameters}. For saturated pixels, the persistence current plateaus at around 0.25 $e^-/$s (Fig.~\ref{fig:h4rg-lo}), leading to persistence signals of several dozens of electrons in a subsequent 140s exposure as shown in the top four panels of Fig.~\ref{fig:pers_joint}. 
The image persistence due to bright stars is comparable in brightness to images of $J=22-24$ galaxies (bottom two panels of Fig.~\ref{fig:pers_joint}). When the bright star persistence signal overlaps the light profile of fainter galaxies, the shapes of the fainter galaxies could be greatly affected. These persistence images exhibit the features of star images such as diffraction spikes. If a galaxy image happens to overlap the diffraction spike of the persistence signal from a bright star, it could introduce directional changes to the galaxy shape, as shown in Fig.~\ref{fig:pers+gal}. In Fig.~\ref{fig:pers+gal}, we directly show an image of a magnitude $J=22$ galaxy and the persistence of a magnitude $J=7$ star to demonstrate the relative contribution of the two components. However, the actual impact of this effect in practice will depend on the relative densities of bright stars and faint galaxies, which will determine the degree to which the persistence can induce coherent galaxy shape correlations.

\begin{figure}
	\centering
	\includegraphics[width=\columnwidth]{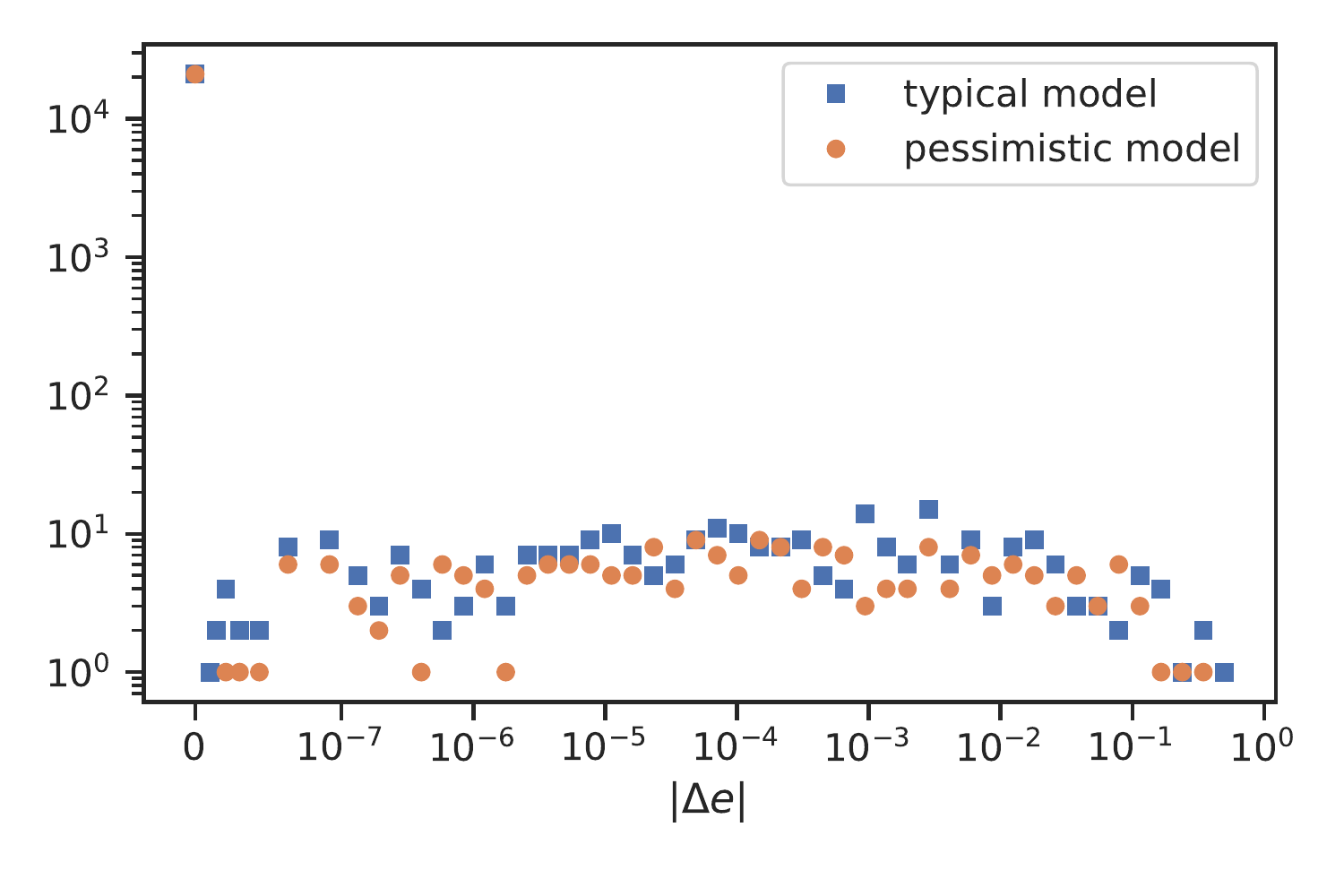}
	\caption{\label{fig:hist_de} Histogram showing the number of galaxies with a given level of shape offset $|\Delta e|$ due to persistence for a subset of 21k simulated galaxies. In this plot, both sets of persistence model parameters are shown in blue and orange as labeled in the legend. Both models have a dot in the upper left corner, indicating that the ellipticities of the majority of galaxies are not affected by persistence. Even though the two models show strong differences in persistence at high signals above 100k, the effects of the two models on galaxy images are similar in practice  because pixels with signal levels greater than 100k e- are quite rare. 
	}
\end{figure}

In Fig.~\ref{fig:hist_de}, we show the distribution of $\Delta e$ values
due to persistence effect using the two sets of persistence model parameters in Table~\ref{tab:parameters}. Note that this plot is based on a smaller subset of 21k simulated galaxies that provides a comparison between the two sets of persistence model parameters. Even though the pessimistic model
has a significantly stronger persistence effect for higher signals above 100k, the two models show similar effects in terms of $\Delta e$. This is because signals greater than 100k electrons are quite rare and only occur at the very center of bright stars. The vast majority of pixels in the images have signals below this value, where the difference in persistence effect between the two sets of model parameters is not significant.  Fig.~\ref{fig:hist_de} shows that around 0.5\% of galaxies have $\Delta e$ greater than $10^{-4}$. This number provides a basic sense for the impact of persistence on galaxy shapes. Besides the percentage of affected galaxies, average shape differences $\langle e_1 \rangle$, $\langle e_2 \rangle$ and $\langle \Delta e \rangle$ are summarized in Table~\ref{tab:mean_e}.

\begin{table}
\begin{tabular}{llll}
\hline
$\langle e_{1, \text{intrinsic}} \rangle$ & $\langle \Delta e_1 \rangle$ & $\langle e_{2, \text{intrinsic}} \rangle$ & $\langle \Delta e_2 \rangle$ \\ \hline
$8.2 \times 10^{-5}$               & $-4.8 \times 10^{-6}$        & $3.6 \times 10^{-4}$               & $-2.3 \times 10^{-6}$    \\ \hline   
\end{tabular}
\caption{\label{tab:mean_e} The mean value of $e_1$, $e_2$ and $\Delta e$ in the simulation containing 0.9 million galaxies.}
\end{table}

\subsection{Correlation between shape errors of galaxies}

\begin{figure}
	\centering
	\includegraphics[width=\columnwidth]{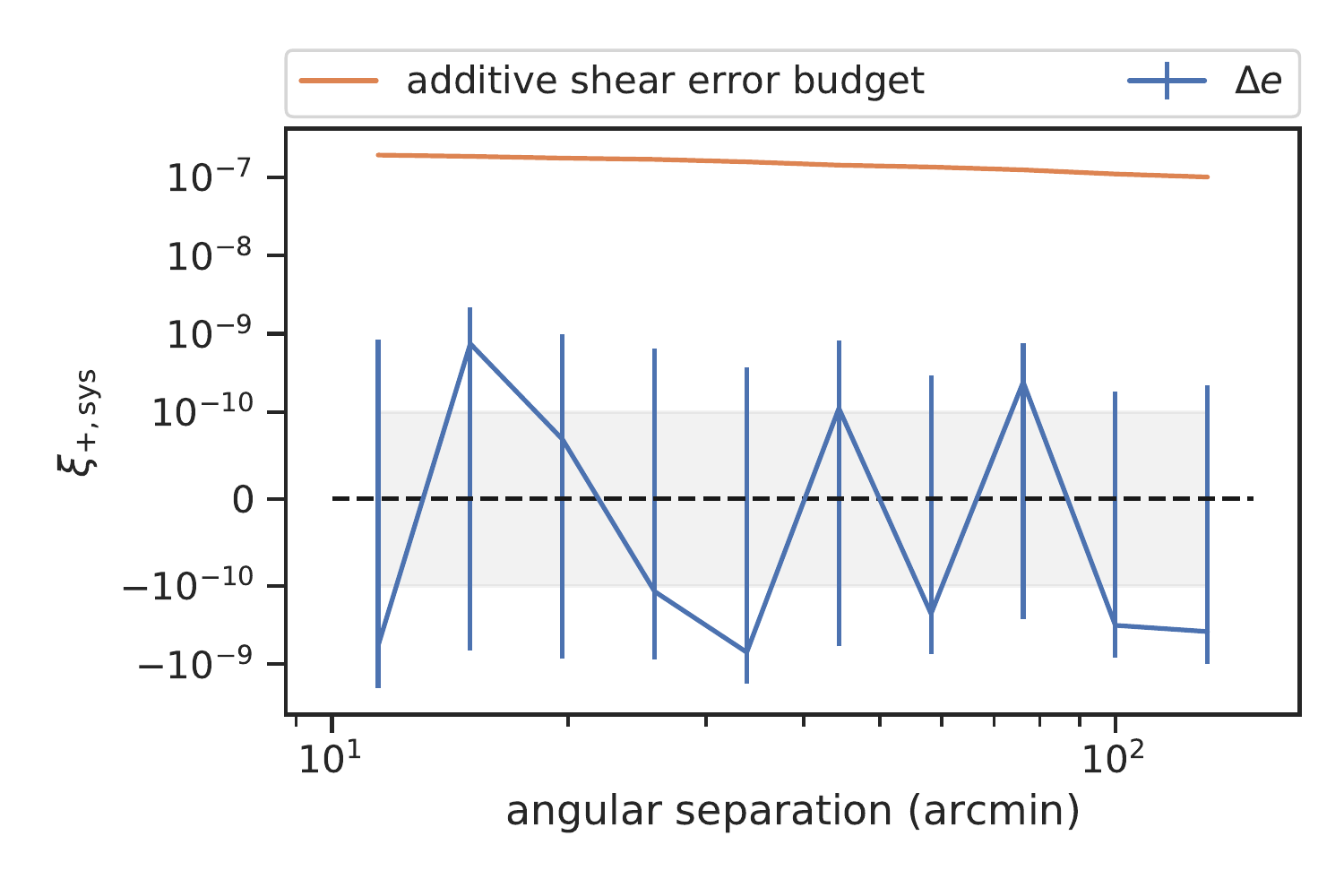}
	\caption{\label{fig:xi} The two-point correlation function $\xi_{+,\text{sys}}$ of $\Delta e$ due to the persistence effect from previous exposures, compared to the total additive shear systematic error budget for the cosmological shear correlation function $\xi_+$ for Roman in symlog y scale (with linear range shaded in grey). The additive shear systematic error is Hankel-transformed from the power spectrum of the Roman High-Latitude Imaging Survey additive shear bias given the requirement in harmonic space. 
	}
	
\end{figure}

In order to quantify the bias of the correlation function $\xi_{\pm}$ of galaxy shapes due to persistence, we measure the correlation function from simulations with and without the persistence effect. Similar to the definition of the shear correlation function measured in weak lensing data and used to constrain cosmological parameters, Eq.~\eqref{eq:xi}, here we define the correlation function $\xi_{\pm}$ based on $\Delta e$ as 
\begin{equation}
    \xi_{\pm,\text{sys}} = \left\langle \Delta e_t \Delta e_t \right\rangle \pm \left\langle \Delta e_\times \Delta e_\times \right\rangle,
\end{equation}
where $\Delta e_t$ and $\Delta e_\times$ are the tangential and the cross ellipticity components relative to the separation vectors of galaxy pairs.
To test whether the persistence effect introduces coherent distortions to galaxy shapes, we measure the correlation of $\Delta e$ due to persistence directly.
In the simulation, a total of 0.9 million galaxies are simulated, with around 1\% of them having $\Delta | e | \ge 1.0\time 10^{-8}$ and around 0.5\% having $\Delta | e | \ge 1.0\time 10^{-4}$.
We use \textsc{TreeCorr}\footnote{\url{https://github.com/rmjarvis/TreeCorr}} to measure the two point correlation functions of ellipticities $\Delta e_1$ and $\Delta e_2$ \citep{2004MNRAS.352..338J}.
In measuring the correlation, all  galaxies are used, regardless of whether they are affected by persistence or not. The result is shown in Fig.~\ref{fig:xi}. The error bars are estimated using the jackknife method with 20 patches over the area. We do not detect significant coherent shifts in ellipticities due to the persistence effect of bright stars. Even though there are cases where strong persistence greatly affect the shapes of galaxies, the average effect on the galaxy ensemble is low. Hence, it is possible to treat the impact of persistence on galaxy shapes as  uncorrelated noise. 
To quantify the significance of the persistence effect on galaxy shapes, we compute the correlation functions of the Roman total additive shear systematic error budget \citep{2018arXiv180403628D, 2021MNRAS.501.2044T} in the same scale as the comparison baseline, as shown in Fig.~\ref{fig:xi}. The level of $\xi_{+,\text{sys}}$ due to persistence is at least one order of magnitude lower than the Roman requirement on the additive shear bias on the angular scales considered, indicating that persistence is a subdominant effect for weak lensing shear compared to more important systematic effects such as photo-z calibration \citep{2019ApJ...877..117H} and detector nonlinearity \citep{2016PASP..128j4001P}.  

\begin{figure}
	\centering
	\includegraphics[width=\columnwidth]{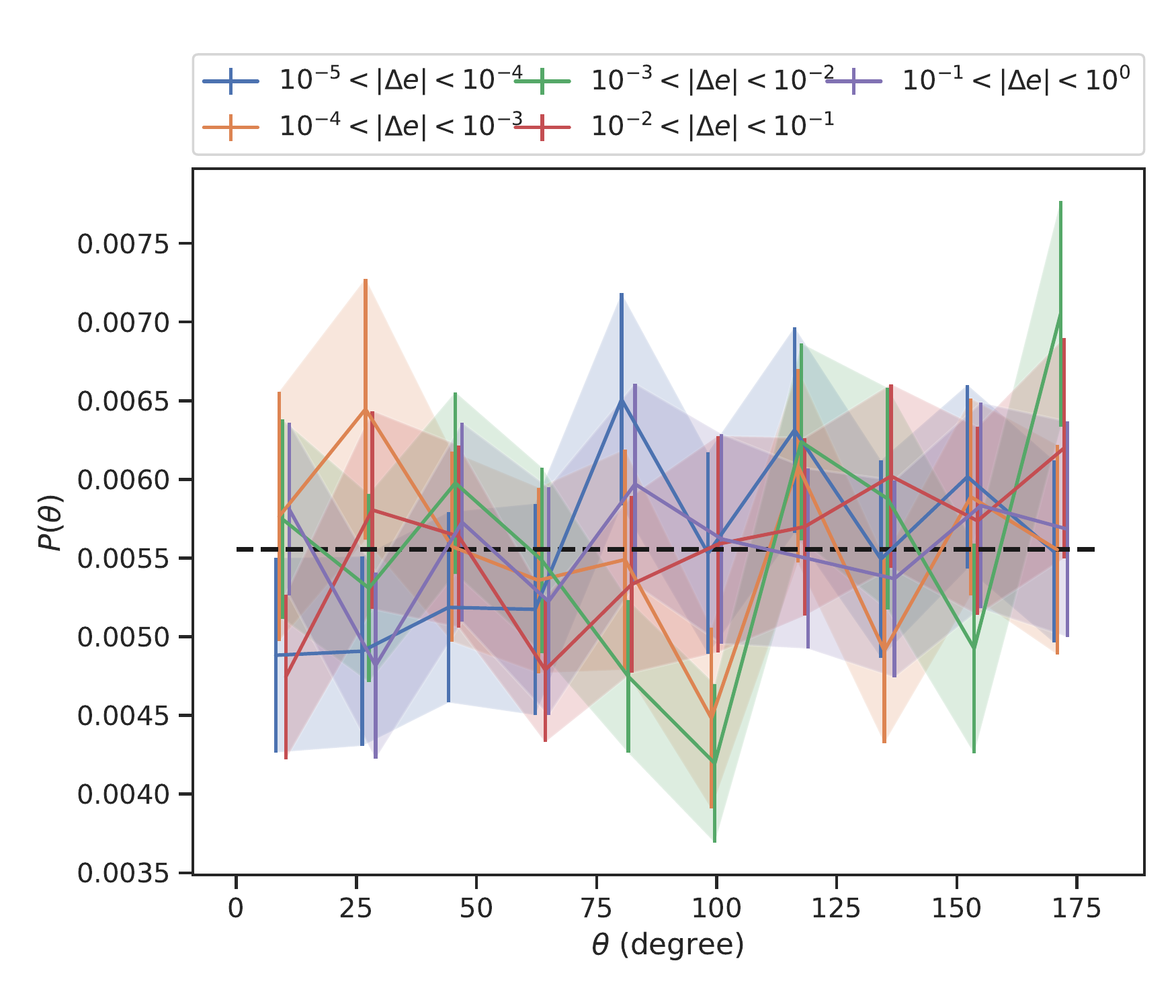}
	\caption{\label{fig:bin_theta} The probability distribution of the position angle $\theta$ of $\Delta e$, which demonstrates the isotropy of $\Delta e$. In this plot, the $\Delta e$ values are binned in 5 bins and the probability is weighted by $\Delta e$ of each galaxy. The flatness of the probability distribution illustrates the  random orientations of the $\Delta e$ values. The error bars and the shaded regions indicate the $1\sigma$ bootstrap uncertainties.  
	}
\end{figure}

To further demonstrate the non-coherent nature of $\Delta e$, we examine probability distributions of the position angle $\theta$ of $\Delta e$ in bins of $|\Delta e|$ and show the results in Fig.~\ref{fig:bin_theta}. The uniform distribution of $\theta$ helps to explain the lack of coherent impact on the galaxy shape correlation function in Fig.~\ref{fig:xi}.

\section{Conclusions}
\label{sec:con}
In this work, we study the impact of image persistence, especially the image persistence of bright stars, on weak lensing signals in the context of the Roman High-Latitude Survey. The images are simulated using the Roman image simulation suite in \cite{2021MNRAS.501.2044T} and the image simulation software \textsc{GalSim} \citep{2015A&C....10..121R}. A total of 0.9~million galaxies are simulated in an area of $6.25~{\rm deg}^2$ with hundreds of dither images. In the simulations, detector effects relevant to Roman Space Telescope are incorporated, but we focus on the persistence effect. An empirical persistence model of the H4RG near-infrared photodiode detector is implemented and applied to model the image persistence of earlier exposures. 

The Roman Project has put significant effort into the design of H4RG flight detector in order to minimize the image persistence. For weak lensing, 
even though no mitigation method for persistence is applied in our simulations to remove the image persistence effect from earlier exposures, we find no significant impact of persistence on galaxy shapes for weak lensing sciences given the current persistence performance of the detectors. The correlation functions of the galaxy shape contamination $\Delta |e|$ due to persistence do not show spatial correlation between the systematic shape biases for the simulated galaxies. A comparison against the Roman total additive shear error budget further indicates that the persistence is a subdominant effect for weak lensing.  

We note that in this work we focus only on the persistence effect due to bright stars. Future work should assess whether there are any  coherent galaxy shape changes due to the persistence effect during the telescope slews.
\section*{Acknowledgments}
The authors acknowledge help and input from Roman study office. We thank the Jeffrey Kruk and Roman detector working group for helps on persistence characteristics. This work is carried out as part of the Roman Science Investigation Team “Cosmology with the High Latitude Survey” supported by NASA grant 15-WFIRST15-0008. CL and MT acknowledge support from NASA under JPL Contract Task 70-711320, “Maximizing Science Exploitation of Simulated Cosmological Survey Data Across Surveys.” CH is additionally supported by the Simons Foundation.

\bibliography{cosmo}

\end{document}